\documentclass[conference]{IEEEtran}
\IEEEoverridecommandlockouts
\usepackage{cite}
\usepackage{amsmath,amssymb,amsfonts}
\usepackage{algorithmic}
\usepackage{graphicx}
\usepackage{textcomp}
\usepackage{xcolor}

\usepackage{algorithm}
\usepackage{caption}
\usepackage{subcaption}
    
\begin{document}

\title{System-Level Evaluation of Beam Hopping in NR-Based LEO Satellite Communication System
}

%\author{\IEEEauthorblockN{Jingwei Zhang,  Dali Qin, Chuili Kong, Feiran Zhao, Rong Li, Jun Wang }
%\IEEEauthorblockA{
%\textit{dept. name of organization (of Aff.)} \\
%Huawei Technologies Co., Ltd., Hangzhou, China \\
%Email: \{zhangjingwei18, qindali, kongchuili, zhaofeiran, lirongone.li, justin.wangjun\}@huawei.com }
%}

%\author{\IEEEauthorblockN{Jingwei Zhang^1,~$Dali Qin$^1$,~Chuili Kong^1,~$Feiran Zhao^1$,~Rong Li^1,~$Jun Wang^1$,~Ye Wang^2}
\author{\IEEEauthorblockN{Jingwei Zhang$^1$, Dali Qin$^1$, Chuili Kong$^1$, Feiran Zhao$^1$, Rong Li$^1$, Jun Wang$^1$, Ye Wang$^2$}
\IEEEauthorblockA{
 $^1$ Huawei Technologies Co., Ltd., Hangzhou, China, Email: zhangjingwei18@huawei.com \\
 $^2$ PengCheng Laboratory, ShenZhen, China, Email: wangy02@pcl.ac.cn
}
}

\maketitle

\begin{abstract}
Satellite communication by leveraging the use of low earth orbit (LEO) satellites is expected to play an essential role in future communication systems through providing ubiquitous and continuous wireless connectivity. This thus has motivated the work in the 3rd generation partnership project (3GPP) to ensure the operation of fifth generation (5G) New Radio (NR) protocols for non-terrestrial network (NTN). In this paper, we consider a NR-based LEO satellite communication system, where satellites equipped with phased array antennas are employed to serve user equipments (UEs) on the ground. To reduce payload weight and meet the time-varying traffic demands of UEs, an efficient beam hopping scheme considering both the traffic demands and inter-beam interference is proposed to jointly schedule beams and satellite transmit power. Then based on NR protocols, we present the first system-level evaluations of beam hopping scheme in LEO satellite system under different operating frequency bands and traffic models. Simulation results indicate that significant performance gains can be achieved by the proposed beam hopping scheme, especially under the distance limit constraint that avoids scheduling adjacent beams simultaneously, as compared to benchmark schemes. 
\end{abstract}


\section{Introduction}

Satellite communication has drawn significant attention in the past decade in both academia and industries owing to its ability to provide ubiquitous wireless coverage and continuous service, especially in areas where the terrestrial coverage of cellular network is not available \cite{9210567}. In terms of orbital height, satellites can be broadly classified into three categories, geostationary earth orbit (GEO) satellites, medium earth orbit (MEO) satellites, and low earth orbit (LEO) satellites. Although supporting smaller coverage compared to GEO and MEO satellites, LEO satellites bring appealing advantages for boosting communication performance, such as reduced over-the-air delay and path loss resulting from the shorter link distance, and lower deployment cost. Therefore, mega-constellation LEO satellite system has emerged as a promising technology in achieving seamless communication service across the globe with fertile business opportunities.

To capitalize the growing opportunities, there has been an upsurge of proposals put forward by different companies about deploying LEO satellite mega-constellations, e.g., OneWeb, Kuiper and Starlink, to provide global broadband access. Besides, the 3rd generation partnership project (3GPP) has joint forces to devote standardization efforts to support the operation of fifth generation (5G) New Radio (NR) in non-terrestrial network (NTN) using satellite access. In Release 15, deployment scenarios, channel models and key impacted areas in NR were identified \cite{V16.1.02020}. Further features and modifications for adapting NR in NTN were studied in Release 16 \cite{V16.1.02021}. In Releases 17 and 18, 3GPP continues the standardization effort to address challenges in the operation of NR-based protocols with NTN, and further investigates the NTN-specific techniques for performance enhancement. 

On the other hand, beam hopping is envisioned as a flexible technique in satellite communication to meet the time-varying traffic demands of user equipment (UEs) on the ground with reduced payload weight and financial cost. It offers a new design degree of freedom for communication performance improvement via intelligently illuminating a portion of the satellite coverage area at each snapshot \cite{5586860}. Specifically, in addition to dynamically adjusting satellite resources, beam hopping can alleviate the inter-beam interference via scheduling beams such that a full bandwidth reuse becomes possible with a higher capacity. Therefore, significant endeavors have been devoted to exploiting and evaluating the benefits brought by beam hopping in satellite communication. In \cite{etsi2005digital}, DVB-S2X standard defined three frame structures with variable frame length and dummy frame, e.g., Format 5-7, to support beam hopping. In \cite{Lauri2017}, the potential of beam hopping was illustrated on top of a system-level simulator, where a GEO satellite is employed to illuminate 14 beams. A cooperative multi-agent deep reinforcement learning framework was proposed in \cite{9693289} for the joint optimization of beam hopping pattern and bandwidth allocation to match the time-varying traffic demand. 

\begin{figure*}[htbp]
\centering
\includegraphics[width=0.72\textwidth, height=5.3cm]{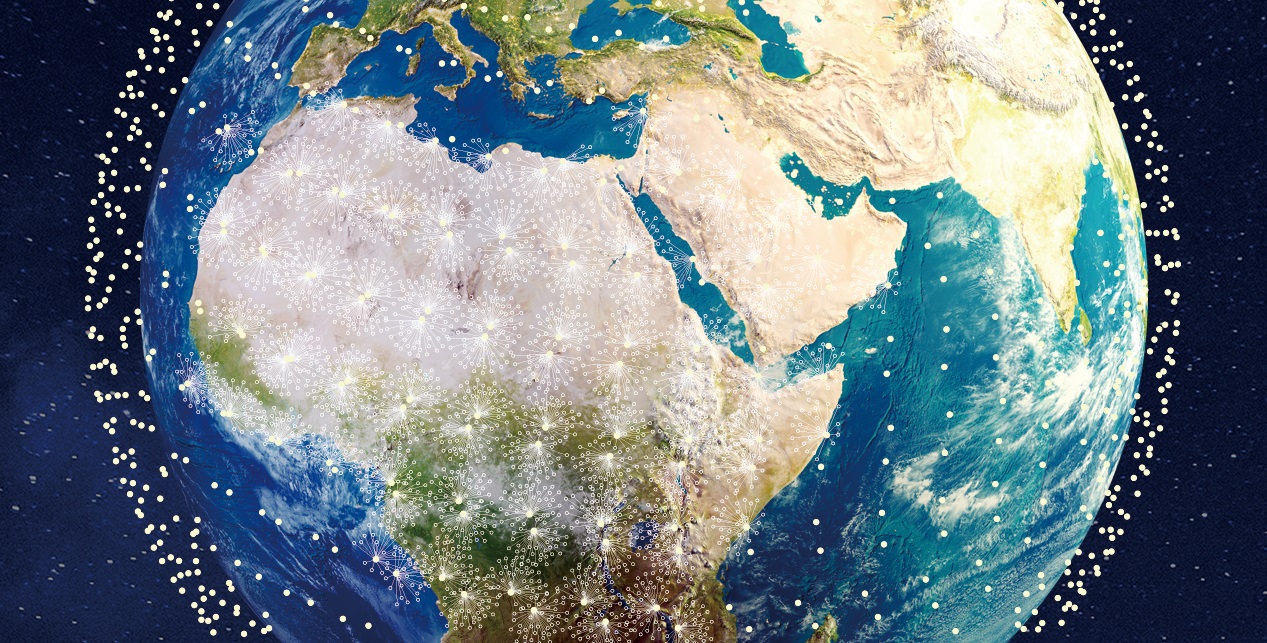}
\caption{Visualization demonstration of one time slot in LEO satellite communication system. \label{snapshot01}}
\vspace{-0.3cm}
\end{figure*}

Motivated by the above, we study in this paper a NR-based mega-constellation LEO satellite communication system, where satellites equipped with phased array antennas are employed to serve UEs on the ground. By focusing on the downlink transmission from satellites to UEs, an efficient beam hopping scheme by taking into account the time-varying traffic demands of UEs and inter-beam interference is proposed to assign beams and allocate transmit power of the satellite. To fully evaluate its performance, we carry out, to the best of our knowledge, the first available system-level evaluations of the beam hopping scheme in a LEO satellite system that utilizes NR protocols. Simulation results show that significant performance gains in terms of throughput, packet life time, and UEs' satisfaction are achieved by our proposed beam hopping scheme.

It is worth noting that different system-level evaluations have been performed for satellite communication systems, but without being fully performed for beam hopping in LEO satellite systems. The authors in \cite{juan20205g} presented the first system-level evaluation of the conventional NR handover algorithm in a LEO satellite system. Thorough system-level simulations were carried out in \cite{9347998} to evaluate the throughput and capacity performance of a non-hopped LEO network.
 
The remainder of this paper is organized as follows. Section \ref{sect01} introduces the simulation methodology used in LEO satellite communication system. Then the layout of spotbeams generated by satellites and an efficient beam hopping scheme are proposed in Section \ref{sect02}. Section \ref{sect03} presents extensive system-level simulation results, followed by conclusions in Section \ref{sect04}.

\section{System-Level Simulation Methodology}
\label{sect01}

In this section, we summarize the system-level simulation methodology in a NR-based LEO satellite communication system, including system overview, antenna configurations as well as relevant protocols, and more details can be found in 3GPP specifications \cite{V16.1.02020} and \cite{V16.1.02021}.

\subsection{System Overview}

This paper considers a LEO satellite communication system, where satellites work with a regenerative payload allowing on-board processing of baseband signals. The system is assumed to operate at two frequency bands for downlink transmission, with a center frequency of 2 GHz and 20 GHz at S band and Ka band, respectively. The system bandwidth is 30 MHz with the subcarrier spacing of 15 kHz for S band, while the bandwidth is 200 MHz and subcarrier spacing is 120 kHz for Ka band. 

Furthermore, the LEO system is assumed to follow a walker constellation, composed of $M$ satellites that are distributed in $P$ planes inclined at $55^{\circ}$, with an orbital altitude equal to 600 km. Each satellite intends to cover the service area via generating at most $K$ spotbeams on the ground, and a maximum number of $I_{\mathrm{max}}\leq K$ beams are assigned at each time slot to illuminate a portion of the coverage area. Details for the designing beam hopping scheme will be elaborated in the next section. Furthermore, a visualization demonstration of one time slot in the considered communication system, which comprises $M=2400$ satellites in $P=60$ planes, for serving parts of the world under $I_{\mathrm{max}} = 40$ is plotted in Fig. \ref{snapshot01}, where satellites and illuminated spotbeams on the ground are represented by yellow points and white circles, respectively. 

\subsection{Antenna Configurations}
\label{antennaconfig}
In the LEO satellite communication system, each satellite is assumed to be equipped with phased array antenna, which can steer multiple beams on the ground and change their directions in real-time. Specifically, the elements of planar array are arranged on a square grid, and the geometry is obtained by placing 28 elements along both the x-axis and y-axis with equal spacing of $0.46\lambda$, where $\lambda$ is a wavelength. The satellite transmit gain $G_T$ is 24 dBi and 30.5 dBi for S band and Ka band, respectively.

By following \cite{V16.1.02021}, the LEO system is assumed to serve handheld UEs and very small aperture terminal (VSAT) UEs in S band and Ka band, respectively. In particular, the handheld UE is equipped with omni-directional antenna with receive antenna gain $G_R=0$ dBi, while the VSAT UE is equipped with a circularly polarized antenna with 60 cm equivalent aperture diameter and receive antenna gain $G_R=39.7$ dBi.

\subsection{Simulator Setup}

%{\color{red}{ more details are needed here}}

The system-level evaluation of the LEO system is performed in a 5G NR simulator with a full protocol stack. For example, radio link control entity is configured in acknowledge mode to support error-free transmission; channel-quality indicator (CQI) information is fed back by UEs for the scheduling of satellite resources; adaptive modulation and coding is employed for the transmission of UEs; retransmissions are processed by adaptive hybrid automatic repeat request (HARQ) protocol. 

%In the simulator, CQI is estimated from the received signal and f

%including RLC acknowledged Mode (AM), , link adaptations with adaptive modulation and coding (AMC) and outer loop link adaptation (OLLA), etc. 

\section{Beam Hopping Methodology}
\label{sect02}

Beam hopping has been shown to be a flexible solution to enhance the communication performance via reducing the payload weight and meeting the time-varying traffic demands. In this section, we describe the beam hopping methodology used in the considered communication system by firstly presenting the method of generating spotbeams for satellites and then proposing an efficient beam hopping scheme.

\subsection{Satellite Spotbeam Layout}

Typically for a UE on the ground, a certain number of satellites are visible above the minimum elevation angle and might provide service to it. In this paper, we assume that satellites employ the earth-fixed beams and each UE can only be served by one satellite at one snapshot such that the satellite can direct its beams towards its associated UEs and interference received from neighboring satellites is reduced. With these assumptions, it is thus essential to find an appropriate association between the layout of satellites' spotbeams and UEs to ensure that beamforming gains are fully exploited. In the following, we propose an efficient approach to find a proper association.
 
After receiving global navigation satellite system (GNSS) signals of UEs, each satellite will assign its spotbeams towards UEs in an order from near to distant until all $K$ spotbeams are utilized or all reporting UEs are covered. Besides, during the assignment, if a UE has already been covered by an existing spotbeam, no new spotbeams will be directed towards it. With the above procedures, UEs are always served by their closest satellites such that they are more likely to enjoy better communication links. Furthermore, we assume a full spectrum reuse across all spotbeams.

\subsection{Beam Hopping Scheme}
\label{beamhopping}

In this subsection, we design a beam hopping scheme to meet the time-varying traffic demands of UEs. Firstly, we denote the satellite set as $\mathcal{M}=\{1,\cdots,M\}$. The corresponding spotbeam set of satellite $m$ is denoted as $\mathcal{K}_m=\{1,\cdots, K_m\}$, with $|\mathcal{K}_m|=K_m\leq K$, $m \in \mathcal{M}$. A set of $\mathcal{J}=\{1,\cdots,J\}$ UEs are distributed requesting communication service. 

For ease of exposition, the time horizon $T$ is equally discretized into $N$ equal time slots, indexed by $n=1,\cdots,N$. The elemental time slot length $\delta=T/N$ generally depends on the scheduling requirement of the LEO communication system, which can be flexibly set as the duration of several slots within a frame or the duration of several frames. The traffic demand of UE $j$ at the beginning of each time slot is denoted as $d_j[n]$, $j\in \mathcal{J}$, which is assumed to vary over time. During each time slot, only a maximum number of $I_{max} \leq K_m$ beams of each satellite are assigned. Let $I_{m,k}[n]$ denote the binary scheduling of beam $k$ of satellite $m$ at time slot $n$, where the corresponding spotbeam is illuminated if $I_{m,k}[n]=1$ and dimmed otherwise. We thus have the following constraints
\begin{align}
\sum_{k=1}^{K_m} I_{m,k}[n] \leq I_{max}, \ \ \forall m \in \mathcal{M}, n=1,\cdots,N,\\
I_{m,k}[n]=\{1,0\}, \ \ \forall k \in \mathcal{K}_m, \forall m \in \mathcal{M}, n=1,\cdots,N.
\end{align}

Denote by $p_{m,k}[n]$ the transmit power for beam $k$ of satellite $m$,  $k \in \mathcal{K}_m$,  $m \in \mathcal{M}$ at time slot $n$. We have the following constraints
\begin{align}
\sum_{k=1}^{K_m} p_{m,k}[n]\leq P_{\mathrm{max}}, \ \  \forall m \in \mathcal{M}, n=1,\cdots,N,
\end{align}
where $P_{{\mathrm{max}}}$ is the maximum transmit power of the satellite. The received power for UE $j$ from its serving satellite can be expressed as
\begin{align}
p_{m,k}^j[n]=p_{m,k}[n]h_{m,k}^j[n],
\end{align}
where $h_{m,k}^j[n]$ is the power gain from satellite $m$ to UE $j$ and is given by 
\begin{align}
h_{m,k}^j[n] =G_t(\theta_{m,k}^j[n], \phi_{m,k}^j[n])+ G_T+G_R  -PL(d_{m,k}^j[n]),
\end{align} 
where $G_t(\theta_{m,k}^j[n], \phi_{m,k}^j[n])$ is the beam gain with $\theta_{m,k}^j[n]$ and $\phi_{m,k}^j[n]$ respectively representing the elevation angle and azimuth angle for UE $j$ at time slot $n$ \cite{9625309}, 
$G_T$ and $G_R$ are satellite transmit gain and UE receive gain, respectively, as described in subsection \ref{antennaconfig}, and $d_{m,k}^{j}[n]$ is the slant path distance between UE $j$ and satellite $m$, and $PL(d_{m,k}^j[n])$ denotes the corresponding path loss as specified in \cite{V16.1.02020}.

As a result, the overall signal-to-interference plus noise ratio (SINR) of UE $j$ at time slot $n$ is calculated as
\begin{align}
\label{sinr}
\gamma_{m,k}^{j}[n]=\frac{I_{m,k}[n] p_{m,k}[n] h_{m,k}^j[n]}{N_0+ \hat{I}^j_{m,k}[n] +\check{I}^j_{m,k}[n]},
\end{align}
where $N_0$ is the noise power determined by noise figure and antenna temperature, $\hat{I}^j_{m,k}[n] $ and $\check{I}^j_{m,k}[n]$ are the intra-satellite interference from beams currently transmitted by satellite $m$ and the inter-satellite interference from beams transmitted by satellite $m' \neq m$ in $\mathcal{M}$, respectively, and are given by 
\begin{align}
\hat{I}^j_{m,k}[n]\triangleq  \sum_{k'=1,k'\neq k}^{K_m}  I_{m,k'}[n]  p_{m,k'}[n] h_{m,k'}^j[n],~~~~~~~~~\!~\\
\check{I}^j_{m,k}[n]\triangleq  \sum_{m'=1, m'\neq m}^M \sum_{k'=1}^{K_{m'}} I_{m',k'}[n]  p_{m',k'}[n] h_{m',k'}^j[n].
\end{align}

It is observed from \eqref{sinr} that designing the beam hopping scheme is in general challenging as it entails high computational complexity with both integer variables $\{I_{m,k}[n]\}$ and non-integer variables $\{p_{m,k}[n]\}$ involved. It can also be found that illuminated spotbeams on the ground should be geographically far from each other to reduce interference so as to increase SINR and thus the offered throughput. Motivated by the aforementioned discussion, we propose an efficient greedy algorithm for the beam hopping scheme while considering the interference and time-varying traffic demands of UEs.

Specifically, at the beginning of each time slot for each satellite, define an empty set as $\mathcal{I}_{m}[n]$ to contain beams that will be assigned, i.e., $\mathcal{I}_{m}[n] \triangleq \{k | I_{m,k}[n]=1\}$. In each iteration, to maintain fairness, a spotbeam $k$ whose maximum UE traffic demand is the largest in $\mathcal{K}_{m}$ will be firstly selected. To reduce the interference, a distance limit constraint is imposed to ensure adjacent spotbeams are not illuminated concurrently. In particular, if and only if the distance $d_{k,k'}$ between the spotbeam $k$ and all other spotbeams $\forall k' \in \mathcal{I}_{m}[n]$ is larger than a pre-specified distance limit $D$, the spotbeam $k$ will be illuminated via setting $I_{m,k}[n]=1$ and moved from set $\mathcal{K}_{m}$ to set $\mathcal{I}_{m}[n]$. Otherwise, the spotbeam $k$ will be directly removed from $\mathcal{K}_{m}$. The process continues until set $\mathcal{K}_{m}$ becomes empty or $|\mathcal{I}_{m}[n]|= I_{\mathrm{max}}$. After determining all illuminated spotbeams, the transmit power of the satellite is evenly distributed among them, i.e., $p_{m,k}[n]=$ $P_{\mathrm{max}}/|\mathcal{I}_{m}[n]|$, $k \in \mathcal{I}_{m}[n]$. The detailed procedures of the proposed beam hopping scheme are summarized in Algorithm \ref{alg01}.

\begin{algorithm}
\caption{Proposed beam hopping scheme. \label{alg01}}
1. Build an empty set $\mathcal{I}_{m}[n]$. \\
2. \textbf{repeat} \\
3. ~~~Select spotbeam $k$ with the maximum UE traffic \\
   \hspace*{0.6cm} demand in $\mathcal{K}_{m}$.   \\
4. ~~~\textbf{if} $d_{k,k'}>D$, $\forall k' \in \mathcal{I}_{m}[n]$.\\
5. ~~~~~~Set $I_{m,k}[n]=1$, and move $k$ from $\mathcal{K}_{m}$ to $\mathcal{I}_{m}[n]$. \\
6. ~~~\textbf{else}\\
7. ~~~~~~Remove $k$ from $\mathcal{K}_{m}$. \\
8. ~~~\textbf{end}\\
9. \textbf{until} $\mathcal{K}_{m}= \emptyset $ or $|\mathcal{I}_{m}[n]|= I_{\mathrm{max}}$.\\
10. Evenly allocate transmit power over all assigned beams,\\ 
\hspace*{0.4cm} $p_{m,k}[n]=P_{\mathrm{max}}/|\mathcal{I}_{m}[n]|$, $k \in \mathcal{I}_{m}[n]$. 
\end{algorithm}

\section{Simulation Results}
\label{sect03}

In this section, extensive simulation results are provided to evaluate the performance of the proposed design. For ease of exposition and to draw the most essential insights, we focus on a specified area in China, where the longitude and latitude span from $103^{\circ}$ to $113^{\circ}$ and from $28^{\circ}$ to $33^{\circ}$, respectively, such that 4 satellites are able to provide services to UEs on the ground simultaneously during the simulation. Each satellite can generate at most $K=100$ spotbeams on the ground with a diameter of 42 km. We consider that 1000 stationary UEs are randomly distributed in the specified area and 881 UEs can be covered by 399 spotbeams of 4 satellites. The maximum transmit power of the satellite is $P_{\mathrm{max}}=300$ W, and the distance limit is $D=42$ km to avoid simultaneous transmission for adjacent or overlapping spotbeams.

For comparison, we consider three beam hopping schemes, namely, 1) beam hopping scheme w/ distance limit, as proposed in Section \ref{beamhopping}; 2) beam hopping scheme w/o distance limit, where the distance limit in the proposed scheme is omitted; 3) round robin scheme, where beams of the satellite are assigned in a round robin pattern. Besides, two traffic models are considered to evaluate the performance of proposed schemes.

\subsection{Full-Buffer Traffic Model}
%To draw useful insights about the system full capacity, 
To draw useful insights about the system full capacity, we firstly consider the full-buffer traffic model, where a new packet of the same size 0.5 Mbytes arrives at the satellite for every UE within each time slot. 

\begin{figure*}[t!]
    %\centering
   %\vspace{-1cm}
    \begin{subfigure}[h]{0.5\textwidth}
        \centering
         %\vspace{-4.2cm}
        %\includegraphics[width=8cm, height=1.1in]{ka_snr_42_02}
        %\includegraphics[width=8cm, height=1.1in]{ka_snr_42_02.pdf}
        %\hspace*{-0.25cm}\includegraphics[width=10cm, height=1.4in]{ka_snr_42_02.pdf}
        \hspace*{-0.6cm}\includegraphics[width=10cm, height=1.4in]{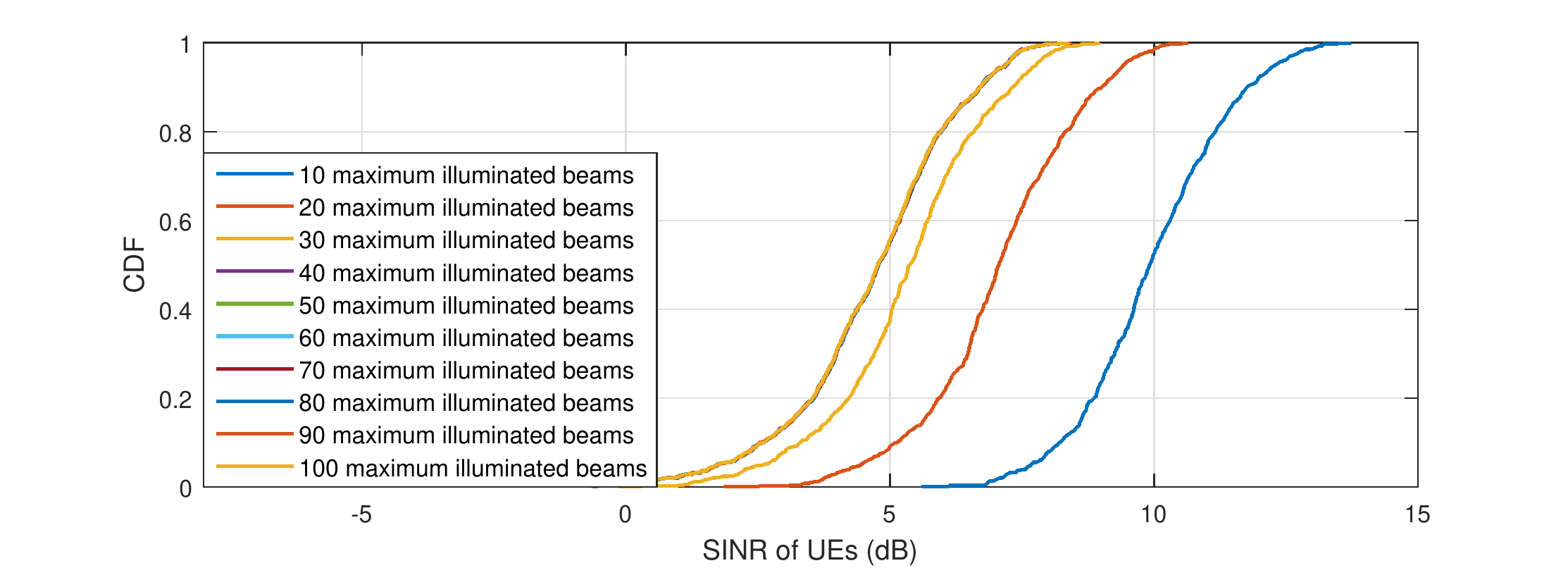}
       % \vspace{-0.2cm}
        \vspace{-0.6cm}
        \caption{SINR of UEs under beam hopping scheme w/ distance limit.}\label{ka_snr_42}
    \end{subfigure}%
    ~ 
    \begin{subfigure}[h]{0.5\textwidth}
        \centering
         \hspace*{-0.3cm} \includegraphics[width=10cm, height=1.4in]{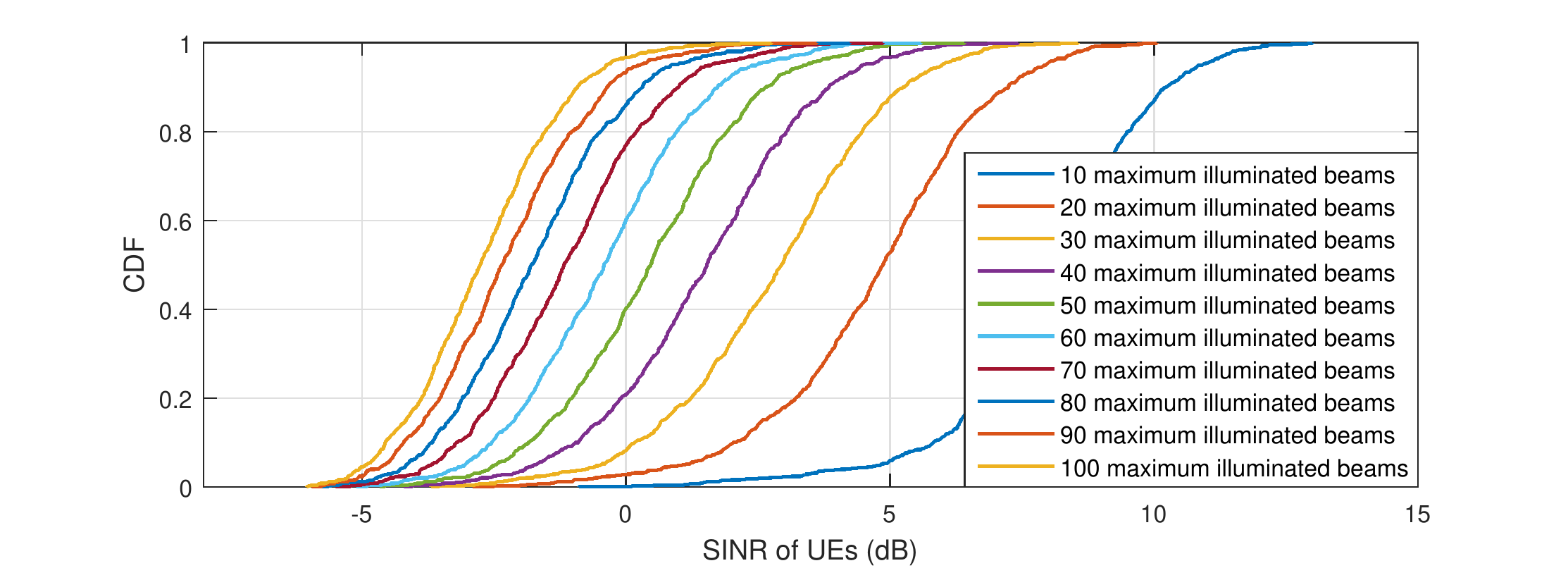}
       %\vspace{-0.2cm}
        \vspace{-0.6cm}
        \caption{SINR of UEs under beam hopping scheme w/o distance limit.\label{ka_snr_1}}
    \end{subfigure} 
% \iffalse
    \begin{subfigure}[h]{0.5\textwidth}
        \centering
        %\vspace{-0.6cm}
        %\includegraphics[width=8cm, height=1.1in]{ka_snr_rr_02.eps}
       % \includegraphics[width=8cm, height=1.1in]{ka_snr_rr_02.pdf}
          %\hspace*{-0.25cm} \includegraphics[width=10cm, height=1.4in]{ka_snr_rr_02.pdf}
          \hspace*{-0.6cm} \includegraphics[width=10cm, height=1.4in]{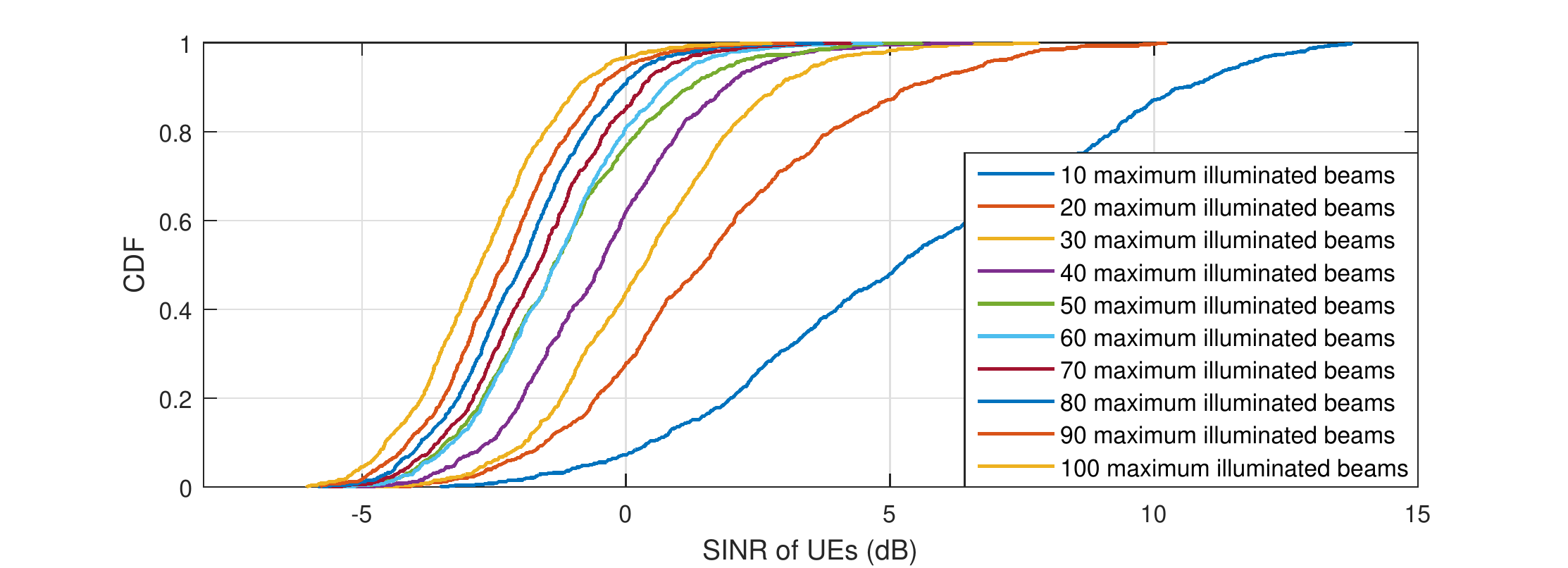}
        % \vspace{-0.2cm}
         \vspace{-0.6cm}
        \caption{SINR of UEs under round robin scheme.\label{ka_snr_rr}}
    \end{subfigure}% 
      ~   
       \begin{subfigure}[h]{0.5\textwidth}
        \centering
        %\vspace{-0.6cm}
        %\includegraphics[width=8cm, height=1.1in]{ka_tp_02.eps}
        %\includegraphics[width=8cm, height=1.1in]{ka-tp-02.pdf}
        \hspace*{-0.3cm} \includegraphics[width=10cm, height=1.4in]{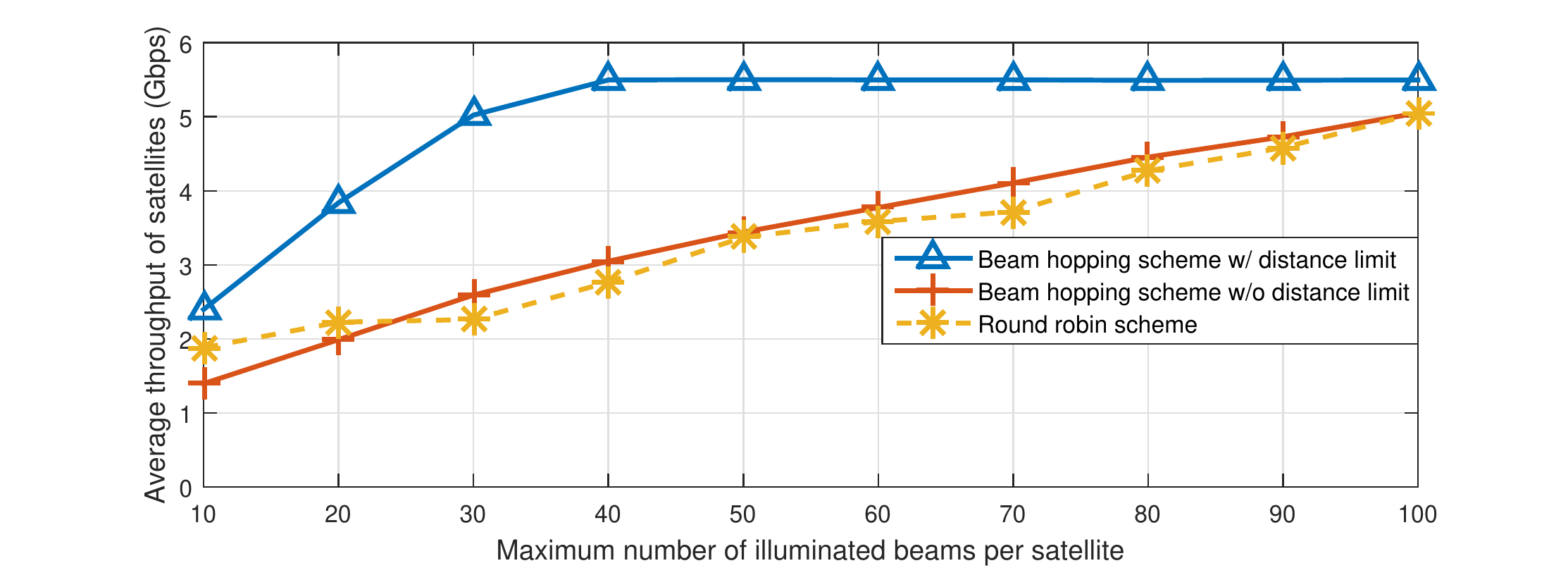}
        %\vspace{-0.2cm}
         \vspace{-0.6cm}
        \caption{Throughput comparison.\label{ka_tp} }
    \end{subfigure}%    
    \vspace{-0.2cm}
    \caption{SINR and throughput comparison between different beam hopping schemes at Ka band. \label{ka}}
\end{figure*}

The downlink SINR and throughput performance with different beam hopping schemes at Ka band are shown in Fig. \ref{ka}. As can be seen, the SINR of UEs gradually decreases with the increase of the maximum number of illuminated beams $I_{\mathrm{max}}$ under the beam hopping scheme w/o distance limit and the round robin scheme in Figs. \ref{ka}(\subref{ka_snr_1}) and \ref{ka}(\subref{ka_snr_rr}), whereas SINR only decreases when the value of $I_{\mathrm{max}}$ is small and almost remains unchanged thereafter for the beam hopping scheme w/ distance limit in Fig. \ref{ka}(\subref{ka_snr_42}). This is expected, since with the increasing value of $I_{\mathrm{max}}$, the transmit power of each beam is decreased due to a given total satellite power budget, thus leading to the degradation of SINR. Whereas for the beam hopping scheme w/ distance limit, where spotbeams within distance $D$ should not be illuminated simultaneously, the number of illuminated beams at each time slot almost remains unchanged even when $I_{\mathrm{max}}\geq 40$, resulting in a similar performance of SINR. Moreover, it is found that SINR of UEs under the beam hopping scheme w/ distance limit is always higher than that under the other two beam hopping schemes, since the inter-beam interference is greatly alleviated. 

\begin{figure*}[t!]
    %\centering
    %\vspace{-1cm}  
     \vspace{-0.3cm} 
    \begin{subfigure}[h]{0.5\textwidth}
        \centering
        %\includegraphics[width=8cm, height=1.1in]{s_snr_42_02.eps}
        %\includegraphics[width=8cm, height=1.1in]{s_snr_42_02.pdf}
      % \hspace*{-0.25cm} \includegraphics[width=10cm, height=1.4in]{s_snr_42_02.pdf}
      \hspace*{-0.6cm} \includegraphics[width=10cm, height=1.4in]{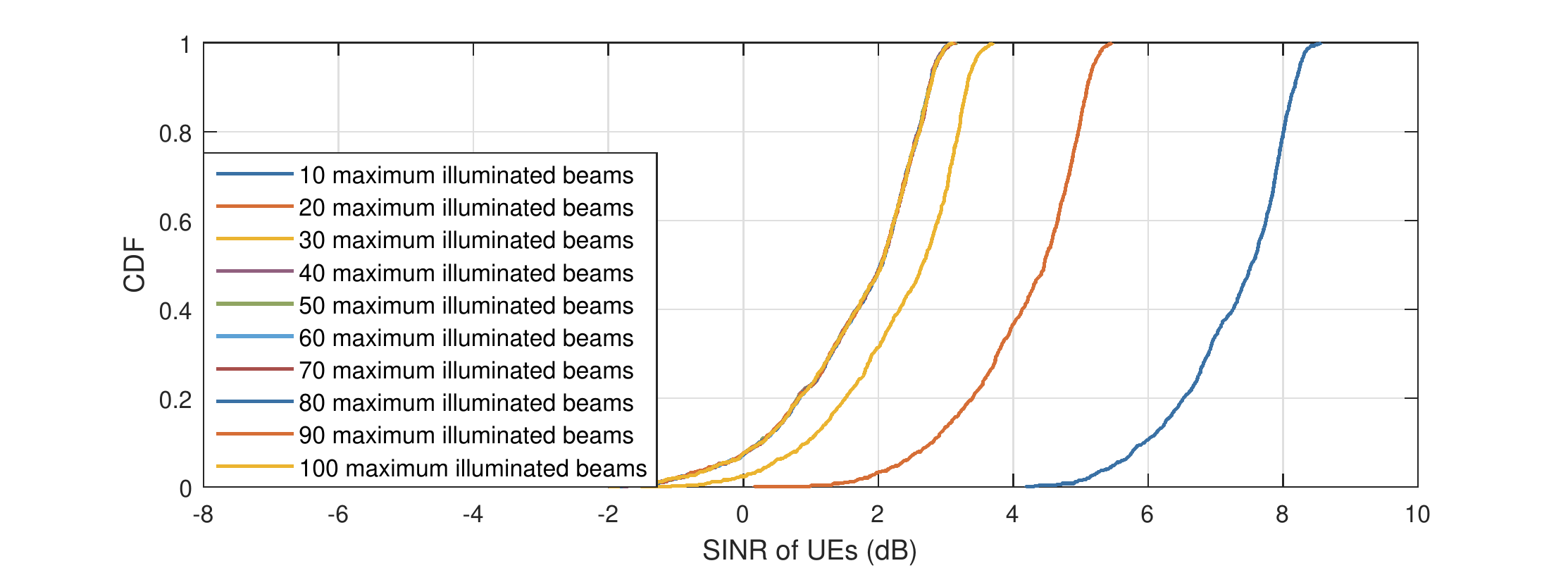}
      % \vspace{-0.2cm}
        \vspace{-0.6cm}
        \caption{SINR of UEs under beam hopping scheme w/ distance limit.\label{s_snr_42}}
    \end{subfigure}%
    ~ 
    \begin{subfigure}[h]{0.5\textwidth}
        \centering
        %\vspace{-0.3cm}
        %\includegraphics[width=8cm, height=1.1in]{s_snr_1_02.eps}
       % \includegraphics[width=8cm, height=1.1in]{s_snr_1_02.pdf}
       \hspace*{-0.25cm}\includegraphics[width=10cm, height=1.4in]{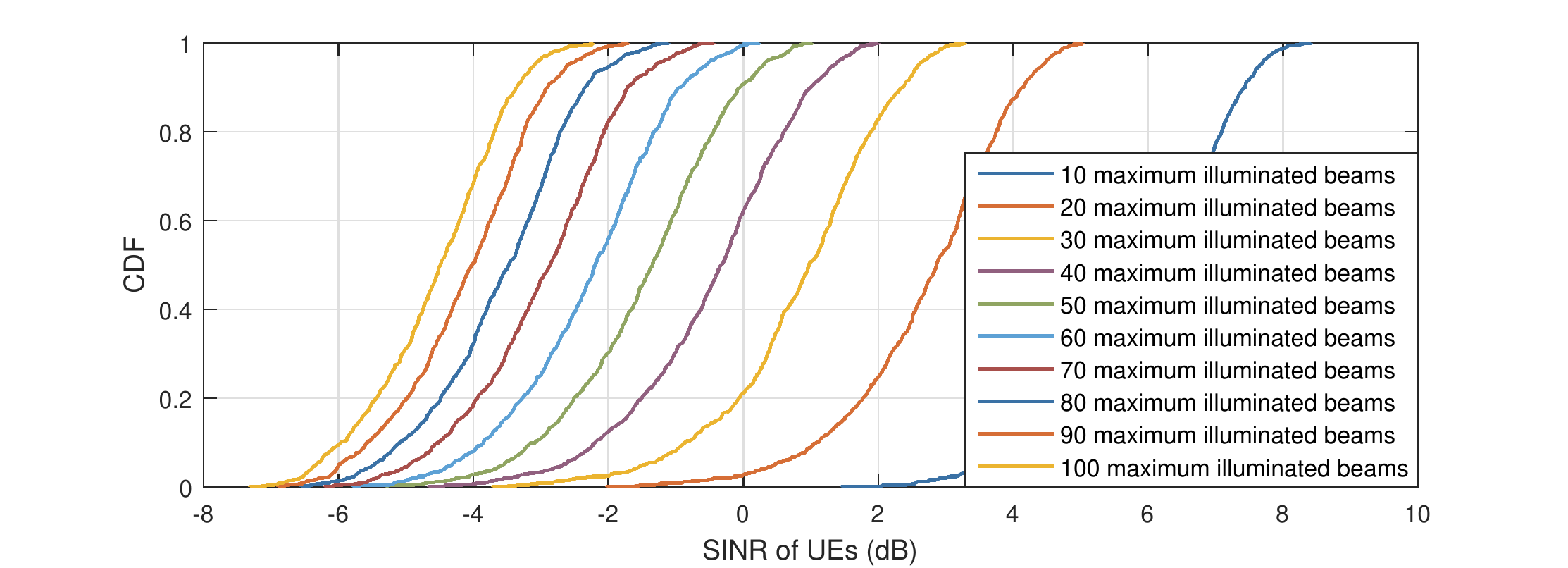}
      %\vspace{-0.2cm}
       \vspace{-0.6cm}
        \caption{SINR of UEs under beam hopping scheme w/o distance limit.\label{s_snr_1}}
    \end{subfigure} 
% \iffalse
    \begin{subfigure}[h]{0.5\textwidth}
        \centering
        %\vspace{-0.8cm}
        %\includegraphics[width=8cm, height=1.1in]{s_snr_rr_02.eps}
        %\includegraphics[width=8cm, height=1.1in]{s_snr_rr_02.pdf}
        % \hspace*{-0.25cm} \includegraphics[width=10cm, height=1.4in]{s_snr_rr_02.pdf}
         \hspace*{-0.6cm} \includegraphics[width=10cm, height=1.4in]{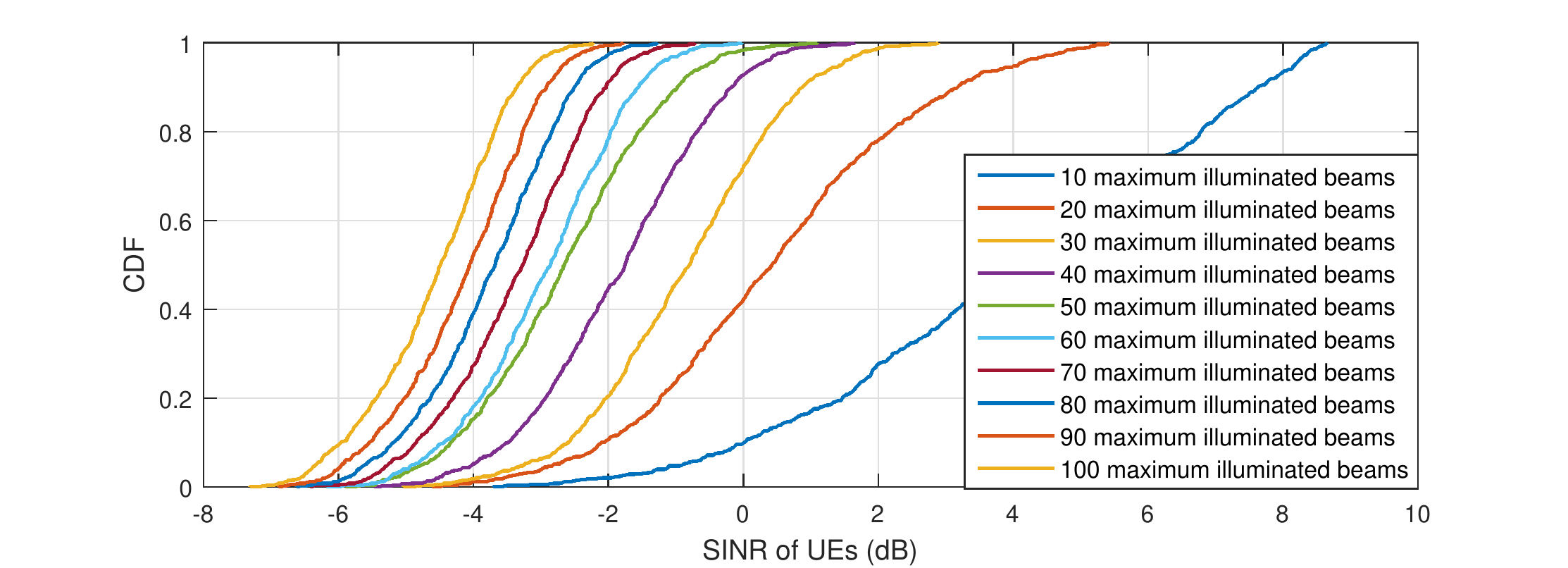}
       % \vspace{-0.2cm}
         \vspace{-0.6cm}
        \caption{SINR of UEs under round robin scheme.\label{s_snr_rr}}
    \end{subfigure}% 
      ~   
       \begin{subfigure}[h]{0.5\textwidth}
        \centering
        %\vspace{-0.8cm}
        %\includegraphics[width=8cm, height=1.1in]{s_tp}
        %\includegraphics[width=8cm, height=1.1in]{s_tp.pdf}
         \hspace*{-0.3cm} \includegraphics[width=10cm, height=1.4in]{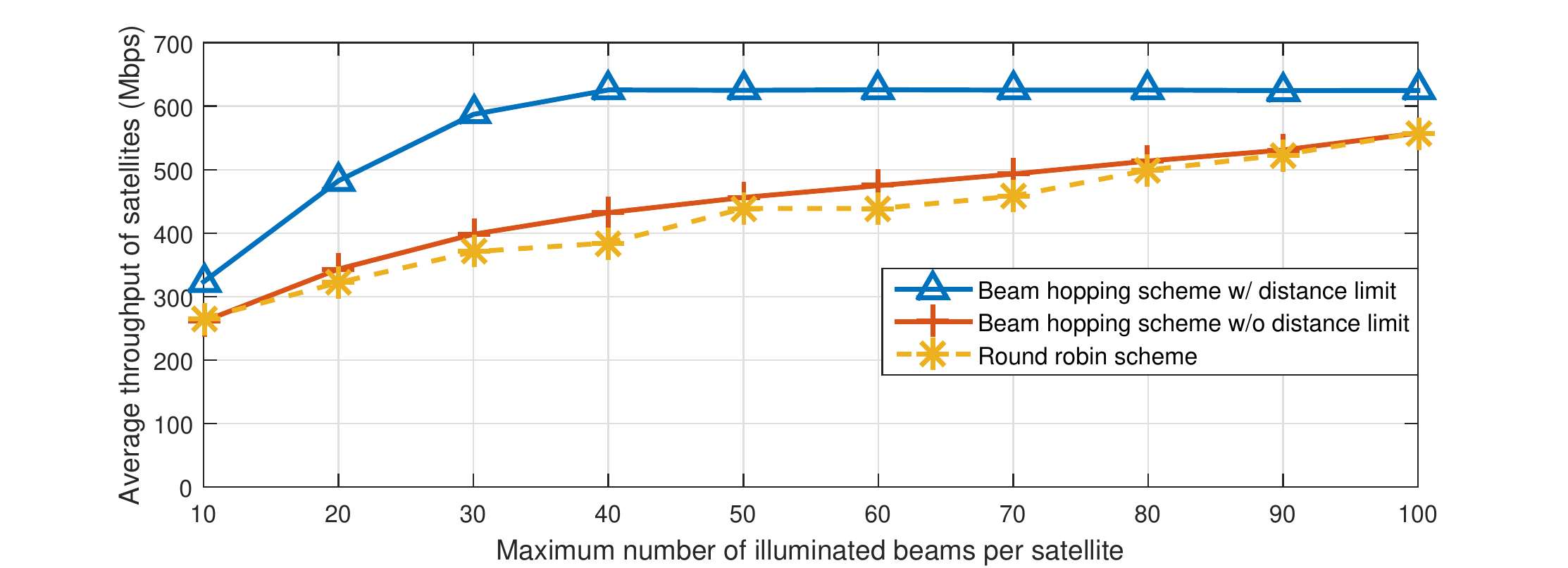}
       % \vspace{-0.2cm}
        \vspace{-0.6cm}
        \caption{Throughput comparison. \label{s_tp} }
    \end{subfigure}%    
    \vspace{-0.2cm}
    \caption{SINR and throughput comparison between different beam hopping schemes at S band. \label{s}}
   \vspace{-0.1cm}
\end{figure*}

The average throughput of each satellite for different maximum number of illuminated beams $I_{\mathrm{max}}$ achieved by three beam hopping schemes is compared in Fig. \ref{ka}(\subref{ka_tp}). It is firstly observed that for a small value of $I_{\mathrm{max}}$, the average satellite throughput increases for all three beam hopping schemes with $I_{\mathrm{max}}$. This is expected, since with the increase of $I_{\mathrm{max}}$, more beams are assigned at each time slot such that higher throughput can be achieved. And the beam hopping scheme w/ distance limit always outperforms the other two schemes due to the suppression of inter-beam interference. Moreover, it is noted that when the maximum number of illuminated beams $I_{\mathrm{max}}$ is relatively small, the round robin scheme slightly outperforms the beam hopping scheme w/o distance limit. The reason is that for UEs with insufficient link budgets, fewer bits are transmitted at each time slot and their accumulated traffic demands are generally larger than other UEs. As a result, spotbeams covering these UEs are illuminated more often to achieve fairness under the beam hopping scheme w/o distance limit, thus achieving a lower average throughput as compared to the round robin scheme.

When the maximum number of illuminated beams $I_{\mathrm{max}}$ is large, it can be observed from Fig. \ref{ka}(\subref{ka_tp}) that the average throughput of satellites under the beam hopping scheme w/ distance limit almost remains constant, due to the distance limit constraints on the number of illuminated spotbeams. Moreover, even when $I_{\mathrm{max}}=100$ and all spotbeams are illuminated at each time slot under the beam hopping scheme w/o distance limit and the round robin scheme, the beam hopping scheme w/ distance limit still outperforms the other two. This further demonstrates effectiveness of the distance limit constraint and illustrates that assigning more beams at each time slot is not necessarily needed in designing the beam hopping scheme.

The downlink SINR and throughput performance with different beam hopping schemes at S band are shown in Fig. \ref{s}. It is observed that UEs in the S band system in general experience a lower SINR than UEs in the Ka band system due to the limited antenna gains. Furthermore, it is noted that the SINR and throughput performance of the three beam hopping schemes at S band follow a similar pattern as the Ka band, and the details are thus omitted for brevity.

\subsection{FTP Traffic Model 3}

\begin{figure*}[t!]
    %\centering
    %\vspace{-1.3cm}
    \begin{subfigure}[h]{0.5\textwidth}
        \centering
        %\includegraphics[width=8cm, height=1.1in]{ka_life_time_n.eps}
        %\includegraphics[width=8cm, height=1.1in]{ka-life-time-n.pdf}
        % \hspace*{-0.25cm}\includegraphics[width=10cm, height=1.4in]{ka-life-time-n.pdf}
        \hspace*{-0.45cm}\includegraphics[width=10cm, height=1.4in]{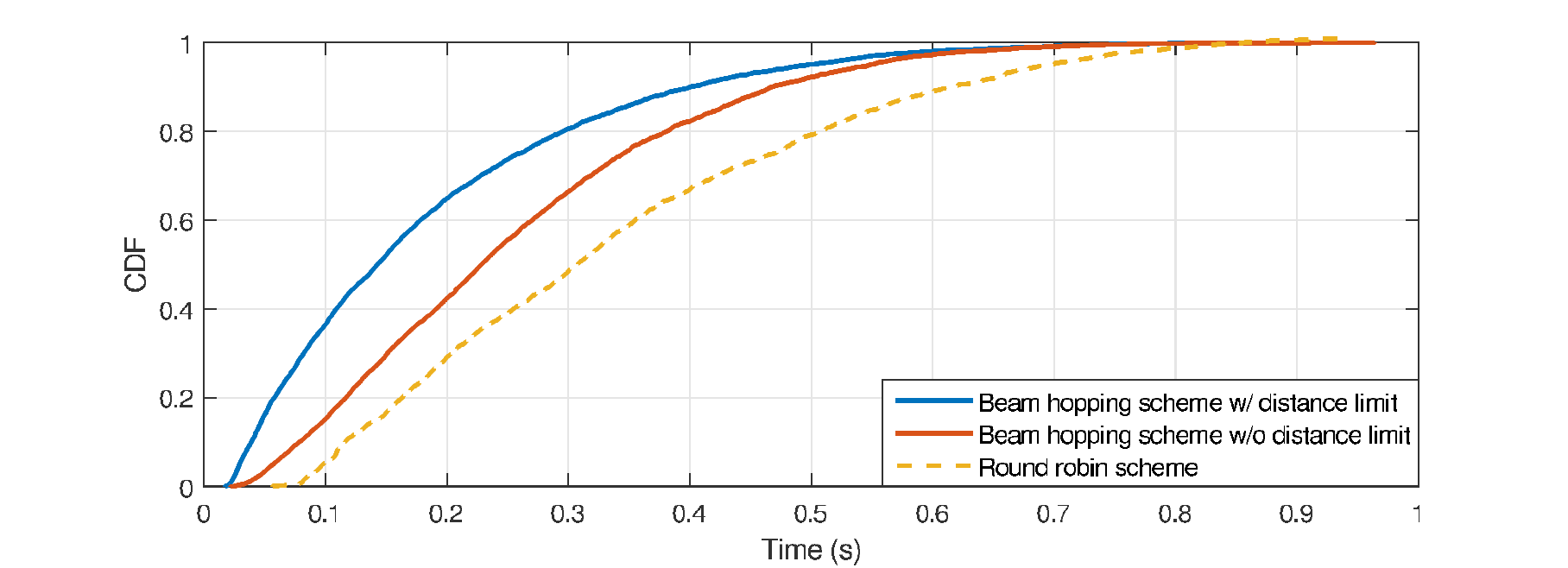}
        %\vspace{-0.2cm}
         \vspace{-0.6cm}
        \caption{Packet life time comparison.}\label{ka_life_time}
    \end{subfigure}%
    ~ 
    \begin{subfigure}[h]{0.5\textwidth}
        \centering
        %\vspace{0.2cm}
        %\includegraphics[width=8cm, height=1.1in]{ka_satis.pdf}
        %\includegraphics[width=8cm, height=1.1in]{ka_satis.eps}
        \hspace*{-0.3cm}\includegraphics[width=10cm, height=1.4in]{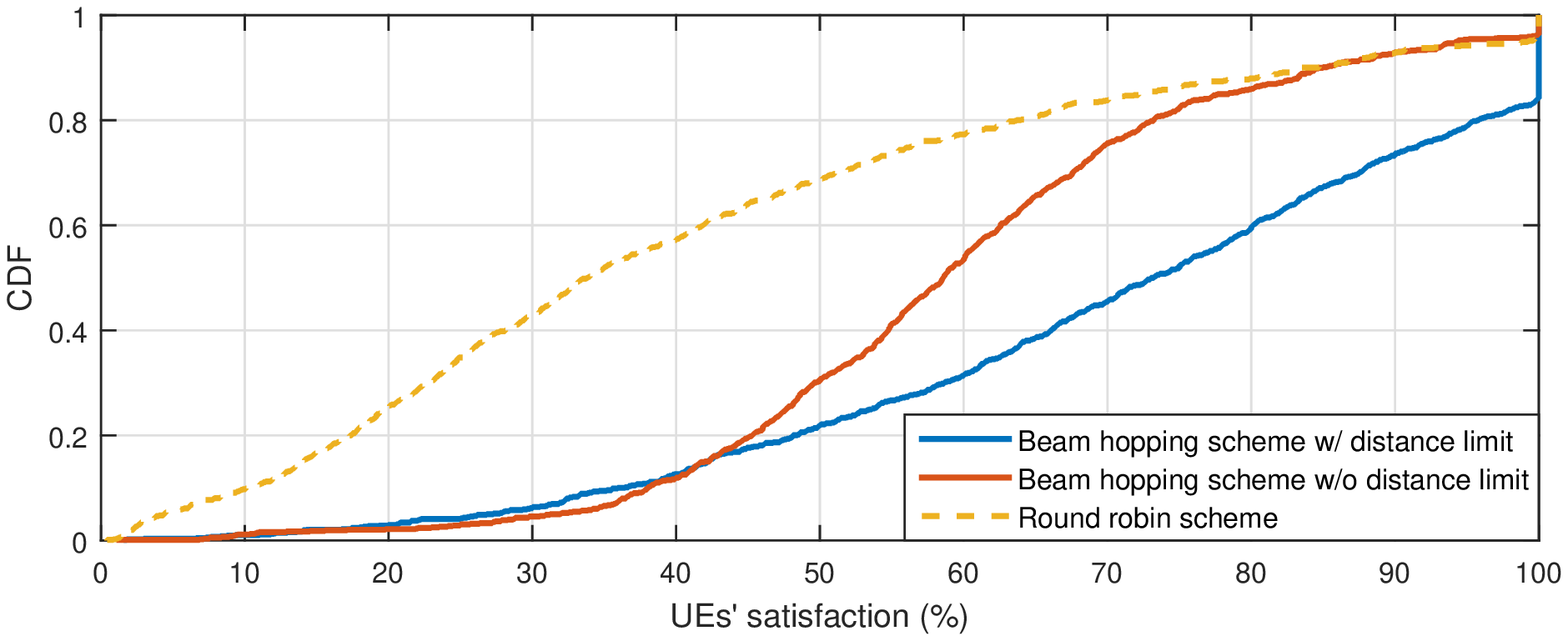}       
        %\vspace{-0.2cm}
        \vspace{-0.6cm}
        \caption{UEs' satisfaction comparison.}\label{ka_satis}
    \end{subfigure} 
    \vspace{-0.2cm}
    \caption{Packet life time and UEs' satisfaction comparison between different beam hopping schemes at Ka band.\label{ka_ftp}}

    \begin{subfigure}[h]{0.5\textwidth}
        \centering
        %\vspace{-0.6cm}
       % \includegraphics[width=8cm, height=1.1in]{s_life_time_ne.jpg}
       %\hspace*{-0.15cm}\includegraphics[width=10cm, height=1.4in]{s_life_time_ne.jpg}
       %\hspace*{-0.35cm}\includegraphics[width=10cm, height=1.4in]{s_life_time_ne.jpg}
       \hspace*{-0.4cm}\includegraphics[width=10cm, height=1.4in]{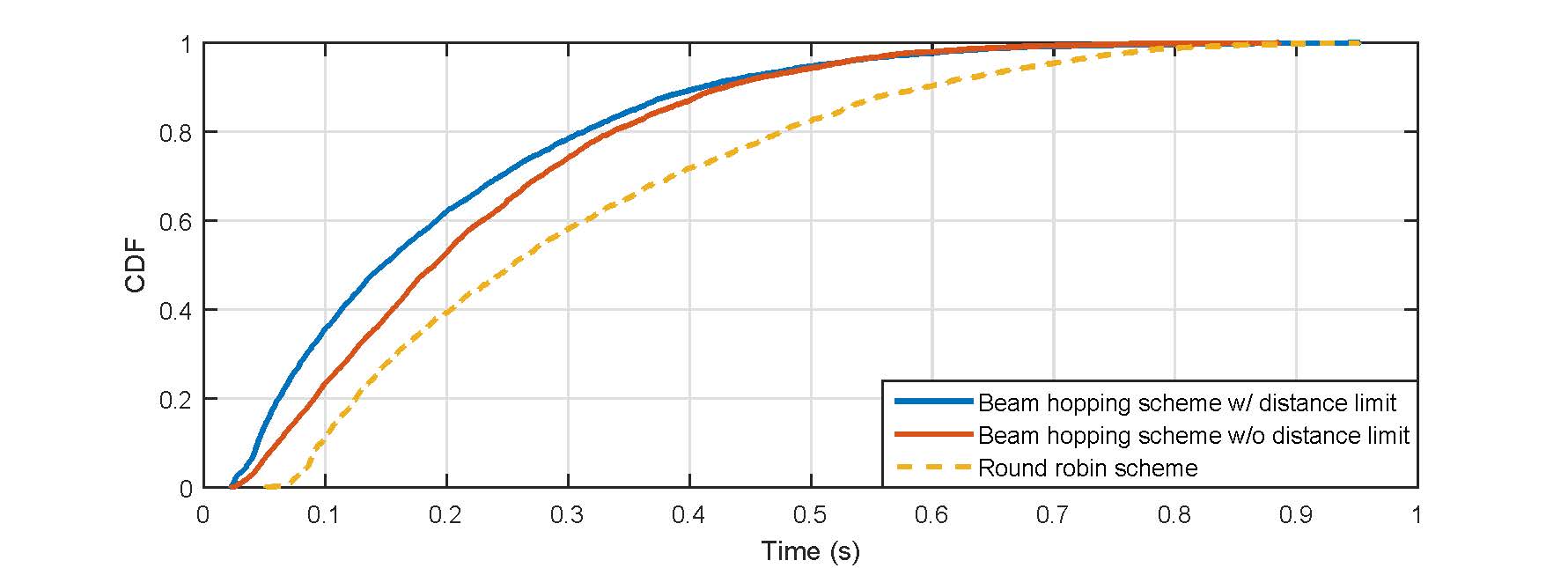}
       % \vspace{-0.2cm}
        \vspace{-0.6cm}
        \caption{Packet life time comparison.}\label{s_life_time}
    \end{subfigure}% 
      ~   
       \begin{subfigure}[h]{0.5\textwidth}
        \centering
        %\vspace{-0.6cm}
        %\includegraphics[width=8cm, height=1.1in]{s_satis.eps}
        %\includegraphics[width=8cm, height=1.1in]{s_satis.pdf}
        \hspace*{-0.3cm}\includegraphics[width=10cm, height=1.4in]{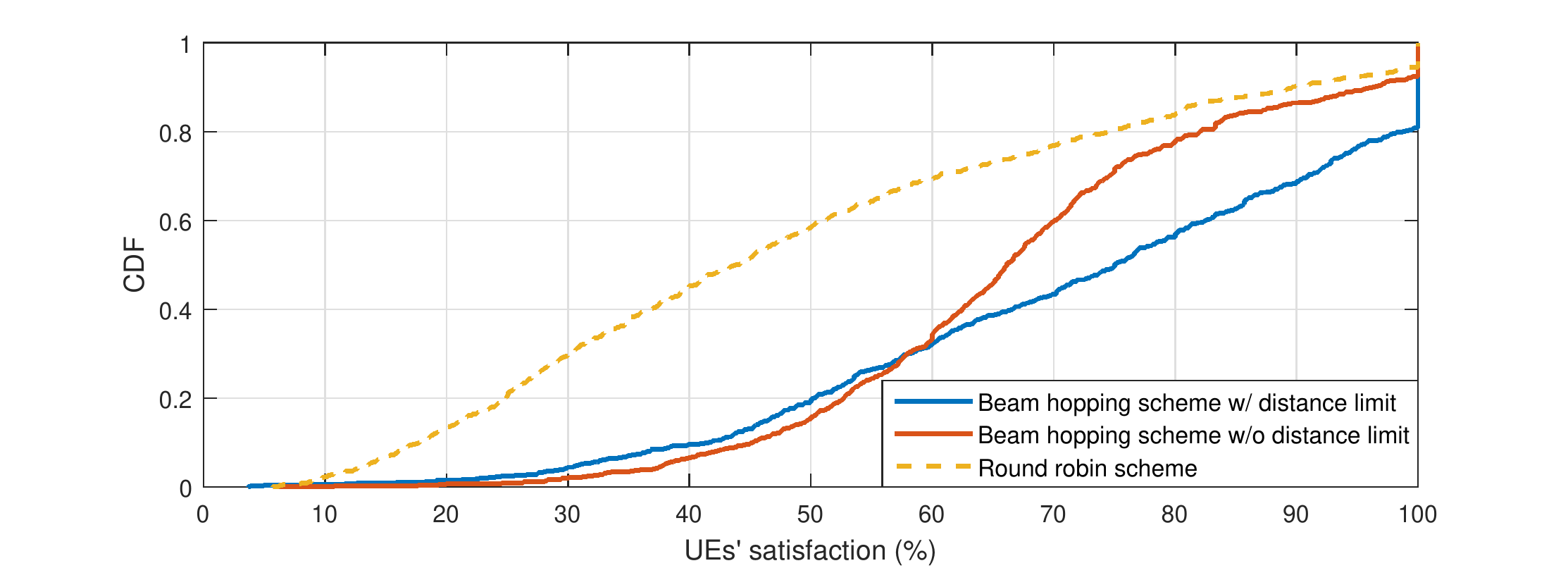}
        %\vspace{-0.2cm}
        \vspace{-0.6cm}
        \caption{UEs' satisfaction comparison. }\label{s_satis}
    \end{subfigure}%
 
    \vspace{-0.2cm}
    \caption{Packet life time and UEs' satisfaction comparison between different beam hopping schemes at S band.\label{s_ftp}}
\end{figure*}

In this subsection, to capture the fact that UEs in the communication system are not always requiring services, a more practical traffic model, file transfer protocol (FTP) model 3 as described in \cite{v36872}, is considered to evaluate the performance of the proposed beam hopping scheme. Packets of sizes 0.5 Mbytes and 0.05 Mbytes are assumed for the Ka band system and S band system, respectively, and they arrive for the same UE according to a Poisson process with arrival rate 8. Two additional performance metrics, which reflect quality of service (QoS) experienced by UEs, are used: packet life time, which is defined as the duration from the time when the packet arrives at satellite to the time when the packet is completely received at UE, and UE's satisfaction, which is defined as the ratio between the offered throughput and demand throughput. The maximum number of illuminated beams is set as $I_{\mathrm{max}}=40$ during the simulation time of $T=1$~s. 

For Ka band, Fig. \ref{ka_ftp}(\subref{ka_life_time}) depicts the life time of all packets completely received at UEs under different beam hopping schemes, while Fig. \ref{ka_ftp}(\subref{ka_satis}) shows the satisfaction of UEs. It is observed that the beam hopping scheme w/ distance limit always outperforms the other two schemes. This further demonstrates the promising benefits of not simultaneously scheduling adjacent spotbeams. For S band, the performance comparison between different beam hopping schemes in terms of packet life time and UEs' satisfaction are shown in Figs. \ref{s_ftp}(\subref{s_life_time}) and \ref{s_ftp}(\subref{s_satis}), respectively, where the proposed beam hopping scheme w/ distance limit enjoys better performance.

\begin{table}[]
\caption{System satisfaction comparison between different beam hopping schemes.}\label{tab01}
\centering
\begin{tabular}{l l l}
\hline
  & Ka band  & S band \\ \hline
  Beam hopping scheme w/ distance limit & 69.5 $\%$  & 70.1 $\%$ \\
  Beam hopping scheme w/o distance limit  & 59.9 $\%$  & 66.3 $\%$ \\ 
  Round robin scheme  & 37.3 $\%$ & 44.9 $\%$ \\ \hline  
\end{tabular}
\vspace{-0.5cm}
\end{table}

Table \ref{tab01} gives the system satisfaction, achieved by different beam hopping schemes. Similarly, it is observed that most traffic demands are satisfied by the proposed beam hopping scheme.

\section{Conclusion}
\label{sect04}

In this paper, we have studied a LEO satellite communication system, where a beam hopping scheme aiming to meet the time-varying demands of UEs and alleviate the interference is proposed to enhance the communication performance via dynamically assigning beams and allocating transmit power of the satellite. Based on the presented simulation methodology and existing NR protocols, this work makes the first attempt to evaluate the performance of the beam hopping scheme in a LEO satellite communication system via system-level simulations. Results show that the throughput performance of different beam hopping schemes in general increases with the maximum number of illuminated beams, but almost remains constant for the beam hopping scheme with distance limit when the maximum number of illuminated beams is large. And the best achievable performance, in terms of throughput, packet life time as well as UEs' satisfaction, is always obtained by the proposed beam hopping scheme with distance limit.

\bibliographystyle{ieeetr}
\bibliography{citing}

\begin{thebibliography}{10}

\bibitem{9210567}
O.~Kodheli and {\emph{et al}.}, ``Satellite communications in the new space
  era: A survey and future challenges,'' {\em IEEE Commun. Surveys Tuts.},
  vol.~23, no.~1, pp.~70--109, 1st Quart., 2021.

\bibitem{V16.1.02020}
{3GPP TR 38.811 V15.4.0}, ``{S}tudy on new radio ({NR}) to support
  nonterrestrial networks,'' {\em Release 15}, Sept. 2020.

\bibitem{V16.1.02021}
{3GPP TR 38.821 V16.1.0}, ``Solutions for {NR} to support non-terrestrial
  networks ({NTN}),'' {\em Release 16}, June 2021.

\bibitem{5586860}
J.~Anzalchi and {\emph{et al}.}, ``Beam hopping in multi-beam broadband
  satellite systems: System simulation and performance comparison with
  non-hopped systems,'' in {\em Proc. Advan. Satell. Multim. Systems Conf. and
  11th Signal Process. Space Commun. Workshop}, pp.~248--255, Sept. 2010.

\bibitem{etsi2005digital}
 {\em Digital video broadcasting ({DVB}); {S}econd Generation Framing
  Structure, Channel Coding and Modulation Systems for Broadcasting,
  Interactive Services, News Gathering and Other Broadband Satellite
  Applications; {P}art 2: {DVB-S2} {E}xtensions (DVB-S2X)}, ETSI Std., Jul.
  2021.

\bibitem{Lauri2017}
L.~Sormunen, J.~Puttonen, and J.~Kurjenniemi, ``System level modeling of beam
  hopping for multi-spot beam satellite systems,'' {\em in Proc. Ka and
  Broadband Commun. Conf. (KaConf)}, Oct. 2017.

\bibitem{9693289}
Z.~Lin, Z.~Ni, L.~Kuang, C.~Jiang, and Z.~Huang, ``Dynamic beam pattern and
  bandwidth allocation based on multi-agent deep reinforcement learning for
  beam hopping satellite systems,'' {\em IEEE Trans. Veh. Technol.}, pp.~1--15,
  Jan. 2022.

\bibitem{juan20205g}
E.~Juan, M.~Lauridsen, J.~Wigard, and P.~E. Mogensen, ``5{G} new radio mobility
  performance in {LEO}-based non-terrestrial networks,'' in {\em Proc. IEEE
  Global Commun. Conf. (GLOBECOM) Workshop}, pp.~1--6, Dec. 2020.

\bibitem{9347998}
J.~Sedin, L.~Feltrin, and X.~Lin, ``{T}hroughput and capacity evaluation of
  5{G} new radio non-terrestrial networks with {LEO} satellites,'' {\em in
  Proc. IEEE Global Commun. Conf. (GLOBECOM)}, pp.~1--6, Dec. 2020.

\bibitem{9625309}
F.~Zhao, Y.~Chen, R.~Li, and J.~Wang, ``On the beamforming of {LEO} earth fixed
  cells,'' in {\em Proc. IEEE Veh. Technol. Conf. (VTC)}, pp.~1--5, Sept. 2021.

\bibitem{v36872}
{3GPP TR 36.872 V12.1.0}, ``Small cell enhancements for {E-UTRA} and {E-UTRAN}
  -- {P}hysical layer aspects,'' {\em Release 12}, 2013.

\end{thebibliography}

\end{document}